\documentclass[aps,prb,twocolumn,superscriptaddress,print]{revtex4}
\usepackage{amssymb,amsmath}
\usepackage{graphicx}
\usepackage{multirow}
\usepackage{hyperref}
\usepackage{appendix}
\usepackage{mathtools}
\usepackage{amsmath}
\usepackage{verbatim}
\usepackage{gensymb}
\usepackage{subfigure}
\usepackage{bm}
\usepackage{threeparttable}
\usepackage{booktabs}
\usepackage{braket}
\usepackage{xcolor}

\begin{document}
\title{Dodecagonal bilayer graphene quasicrystal and its approximants}
\author{Guodong Yu}
\affiliation{Key Laboratory of Artificial Micro- and Nano-structures of Ministry of Education, and School of Physics and Technology, Wuhan University, Wuhan 430072, China}
\affiliation{Institute for Molecules and Materials, Radboud University, Heijendaalseweg 135, NL-6525 AJ Nijmegen, Netherlands}
\author{Zewen Wu}
\affiliation{Key Laboratory of Artificial Micro- and Nano-structures of Ministry of Education, and School of Physics and Technology, Wuhan University, Wuhan 430072, China}
\author{Zhen Zhan}
\affiliation{Key Laboratory of Artificial Micro- and Nano-structures of Ministry of Education, and School of Physics and Technology, Wuhan University, Wuhan 430072, China}
\author{Mikhail I. Katsnelson}
\affiliation{Institute for Molecules and Materials, Radboud University, Heijendaalseweg 135, NL-6525 AJ Nijmegen, Netherlands}
\author{Shengjun Yuan}
\email{s.yuan@whu.edu.cn}
\affiliation{Key Laboratory of Artificial Micro- and Nano-structures of Ministry of Education, and School of Physics and Technology, Wuhan University, Wuhan 430072, China}
\affiliation{Institute for Molecules and Materials, Radboud University, Heijendaalseweg 135, NL-6525 AJ Nijmegen, Netherlands}

\begin{abstract}
Dodecagonal bilayer graphene quasicrystal has 12-fold rotational order but lacks translational symmetry which prevents the application of band theory. In this paper, we study the electronic and optical properties of graphene quasicrystal with large-scale tight-binding calculations involving more than ten million atoms. We propose a series of periodic approximants which reproduce accurately the properties of quasicrystal within a finite unit cell. By utilizing the band-unfolding method on the smallest approximant with only 2702 atoms, the effective band structure of graphene quasicrystal is derived. Novel features, such as the emergence of new Dirac points (especially the mirrored ones), the band gap at $M$ point and the Fermi velocity are all in agreement with recent experiments. The properties of quasicrystal states are identified in the Landau level spectrum and optical excitations. Importantly, our results show that the lattice mismatch is the dominant factor determining the accuracy of layered approximants. The proposed approximants can be used directly for other layered materials in honeycomb lattice, and the design principles can be applied for any quasi-periodic incommensurate structures.
\end{abstract}

\maketitle  

\section{INTRODUCTION}
The bilayer graphene with van der Waals interlayer interaction shows rich electronic properties that are depending on the stacking order and twist angle\cite{BG_AC,BG_Butterfly,BG_ES_0,BG_ES_theta,BG_FlatBand,BG_FractalLL,BG_MoireBand,BG_Mott_Superconducting,BG_OP,BG_QHE,BG_TL,BG_VanHove0,BG_VanHove1,BG_coherence,BG_superconducting}. Two graphene layers can be arranged in AA, AB or twisted configurations. AA and AB stacking are two well-known configurations with electronic properties drastically different from the monolayer.\cite{BG_AA0,BG_AB0,BG_AB1,BG_AB2}. The twist angle between the layers offers an additional degree of freedom to tune the electronic properties. If the commensuration condition is satisfied\cite{commensuration_condition}, namely, the twist angle is $\theta = cos^{-1}({{3q^2-p^2}\over{3q^2+p^2}})$ , where $p$ and $q$ are integers, the tBG forms Moir{\'e} pattern, which is periodic in the space and has an elementary unit cell. The incommensurate tBG, on the other hand, has only quasi-periodicity without any translational symmetry. By now, the tBG with 30{\degree} twist angle, i.e. dodecagonal graphene quasicrystal, which will be referred as graphene quasicrystal for simplicity in the rest of the paper, has been grown successfully on different substrates, including 4H-SiC(0001)\cite{science_QC}, Pt(111)\cite{pnas_QC} and Cu-Ni(111)\cite{cm_QC} surfaces. The interplay between the quasi-periodicity and interlayer interaction results in the emergence of extra Dirac points\cite{science_QC} and critical eigenstates\cite{arXiv_QC}. These novel properties are very different from those of graphene, although the twist angle is even more than 15{\degree}. Some of them have been explained by the k-space tight-binding method proposed in Ref. \cite{arXiv_QC}. However, the calculated quasi-band structure is supercell-like, and can not be compared directly with the ARPES measurements. The overall understanding of the quasi-periodicity on the electronic and optical properties of graphene quasicrystal is still missing. 

In this paper, we study the electronic properties of graphene quasicrystal by using large-scale tight-binding calculations in real space and propose a series of approximants with finite unit cell. For a quasicrystal, its approximant is a periodic structure that contains similar compositions and almost the same local atomic structure within a unit cell. For graphene quasicrystal, although the structure of an approximant can be different from the original quasicrystal, a good approximant should form the same rotational order within a unit cell and keep similar physical properties as the infinite quasicrystal. The approximant method has been widely used to understand the physics of a quasicrystal (for a review see Ref. \cite{approximant_RPM} and references therein). Importantly, theoretical studies on a periodic approximant are much more easier comparing to a real quasicrystal, since methods established for crystalline in condensed matter can be readily used. We are going to set up reliable and convincing approximants for recently fabricated graphene quasicrystal with 12-fold rotational order.

To construct an accurate approximant, one needs first calculate properly the characteristics of a real quasicrystal without any further approximation. This is indeed a quite challenging problem from a computational point of view, as a quasicrystal is not periodic in the space and it contains, in principle, infinite number of sites. In this paper, graphene quasicrystal is modeled by a round disk of exactly 30{\degree} tBG. In order to reproduce the bulk properties of an infinite graphene quasicrystal, such as density of states and optical conductivities, a large enough round disk is needed to get rid of the influence of the edge states. The crucial issue is therefore how to efficiently calculate these properties of the sample with a large radius. It is for sure beyond the commonly used density-functional theory but may be accessible within the tight-binding model. In fact, for twisted bilayer graphene and transitional metal dichalcogenides, there are accurate tight-binding models developed from first principle calculations\cite{fang2015ab,cappelluti2013tight,TB_cite_prb}. The hopping parameters in these models have both distance and orientation dependence and can be used to describe properly interlayer interactions in quasi-periodic bilayer. We will use the one developed in Ref. \cite{TB_cite_prb} which has been verified by comparing results with several experiments \cite{TB_cite_prb,TB_cite_prl,TB_cite_arxiv}. This model has been implemented in our home-made code of tight-binding propagation simulator (Tipsi) and reproduces successfully experimentally observed pseudo-magnetic fields in twisted and relaxed bilayer graphene lattice \cite{TB_cite_arxiv}. 

For a large round disk of graphene quasicrystal described by the tight-binding model, such as a system as large as ten million atoms, the calculations of the electronic properties is still challenging. A common approach one may consider is to diagonalize the Hamiltonian. However, in the diagonalization processes the costs of memory and CPU time scale as $O(N^2)$ and $O(N^3)$, respectively. In view of the potentially large number of arithmetic calculations, it is advisable to use 13–15 digit floating-point arithmetic which corresponds to 8 bytes for a real number and 16 bytes for a complex number. Thus, for a tight-binding sample with $N$ sites, we need in total $16\times N^2$ bytes memory for the storage of all the eigenstates. Considering a common computational node with 256 GB memory, it has a maximum storage for eigenstates of about 126491 sites, which is indeed too small to represent properly a quasicrystal. In fact, our numerical tests show that even a quasicrystal sample with two million sites is not large enough to neglect the influence of the edge states (data not shown). One may use large supercomputers with more nodes and memory to increase the number of sites that can be reached, but it is computational too expensive as the method based on diagonalization does not scale linearly with the size of the sample. We will apply a different approach, tight-binding propagation method (TBPM)\cite{TBPM}, to overcome the difficulties raised by diagonalization. TBPM is based on the numerical solution of time-dependent Schr\"odinger equation without any diagonalization, and importantly, both memory and CPU costs scale linearly with the system size. The calculations of electronic, optical, transport and plasmonic properties can be easily implemented in TBPM without requirement of any symmetry. In this paper, our main purpose is to build approximants and reproduce the electronic properties that have been observed in the experiments. More studies of other properties, which are not presented here, can be further explored by TBPM or applying band theory on these small approximants proposed.

\section{METHODS}\label{method}

\subsection{Effective band structure}
First of all, the spectral function at wavevector $\bm{k}$ and energy $\epsilon$ can be calculated by\cite{BandUnfold_SF}
\begin{equation}
A(\bm{k},\epsilon) = \sum_{I\bm{k}_{SC}}P_{I\bm{k}_{SC}}(\bm{k})\delta(\epsilon-\epsilon_{I\bm{k}_{SC}}),
\end{equation}      
where $\epsilon_{I\bm{k}_{SC}}$ is the energy for $I^{th}$ band at wavevector $\bm{k}_{SC}$ for the approximant. Actually, only one $\bm{k}_{SC}$, namely $\bm{k}_{SC} = \bm{k} + \bm{G}$ being $\bm{G}$ the reciprocal lattice vector of the approximant, contributes to the spectral function. 
The spectral weight is defined by
\begin{equation}
P_{I\bm{k}_{SC}}(\bm{k})=\sum_{s=1}^2 \sum_i \left|  \braket{\psi^{PC_s}_{i\bm{k}}| \Psi^{SC}_{I\bm{k}_{SC}}}  \right|^2=\sum_{s=1}^2 P_{I\bm{k}_{SC}}^s(\bm{k}),
\end{equation}
where $\left| \psi^{PC_s}_{i\bm{k}}\right>$  and $\left|\Psi^{SC}_{I\bm{k}_{SC}} \right>$ are the eigenstates of layer $s$ and the approximant, respectively. Under the tight-binding method, the spectral weight contributed from layer $s$ can be described by
\begin{equation}\label{spec_weig}
P_{I\bm{k}_{SC}}^s(\bm{k})= {{1}\over{n_s}}\sum_\alpha \sum_{\bm{l_s}\bm{l_s}^{'}}e^{i\bm{k}\cdot(\bm{l_s}-\bm{l_s}^{'})}U^{\bm{l_s}\alpha^*}_{I\bm{k}_{SC}} U^{\bm{l_s}^{'}\alpha}_{I\bm{k}_{SC}}.
\end{equation}
Here, $n_s$ is the number of primitive unit cell of layer $s$ in one elementary unit cell of the approximant. $U^{\bm{l_s}\alpha}_{I\bm{k}_{SC}}$ is the projection of $\left|\Psi^{SC}_{I\bm{k}_{SC}} \right>$ (the eigenstate of the approximant) on $\left| \bm{k}_{SC} \bm{l_s}\alpha\right>$ (the Bloch basis function of approximant), which can be constructed by all atomic orbitals
\begin{equation}
\left|\bm{k}_{SC}\bm{l_s}\alpha \right> = {{1}\over{N}}\sum_{\bm{L}} e^{i\bm{k}_{SC}\cdot\bm{L}} \phi(\bm{r}-\bm{L}-\bm{l_s}-\bm{t}_\alpha),
\end{equation}   
where $N$ is the number of the SC. $\phi(\bm{r}-\bm{L}-\bm{l_s}-\bm{t}_\alpha)$ is the $p_z$ orbital located at $\bm{L}+\bm{l_s}+\bm{t}_\alpha$. Equation (\ref{spec_weig}) indicates that only the eigenstates of approximant are necessary to obtain the spectral function. 

Then, the effective band structure can be obtained by\cite{BandUnfold_EBS}
\begin{equation}
\delta N(\bm{k},\epsilon) = \int_{\epsilon-\delta \epsilon/2} ^ {\epsilon+\delta \epsilon/2} A(\bm{k}, \epsilon^{'}) d \epsilon^{'},
\end{equation}
where $\delta \epsilon$ is the bin width in energy sampling.

\subsection{Density of states}
The density of states (DOS) are calculated by TBPM\cite{TBPM} based on the numerical solution of the time-dependent Schr\"odinger equation. In order to calculate the DOS under a magnetic field and then Hofstader's butterfly, the hopping is replaced by the Peierls substitution. In TBPM, a random superposition of the $p_z$ orbitals at all sites is used as the initial state $\left|\phi_0 \right>$ with $\braket{\phi_0|\phi_0}=1$. DOS is calculated as Fourier transform of the time-dependent correlation function
\begin{equation}
d(\epsilon)={{1}\over{2\pi}}\int_{-\infty}^\infty e^{i\epsilon\tau}\braket{\phi_0|e^{-iH\tau/\hbar}|\phi_0}d\tau. 
\end{equation}
During all calculations in TBPM, we always use systems with more than 10 million atoms, for graphene quasicrystal, it is a disk with radius about 204 nm. The open and periodic boundary conditions are used for graphene quasicrystal and the approximants, respectively. 

\subsection{Optical conductivity}
The optical conductivity is calculated by using the Kubo formula in TBPM\cite{TBPM}. The real part of the optical conductivity matrix $\sigma_{\alpha,\beta}$ at temperature $T$ reads
\begin{equation}
\begin{aligned}
Re\sigma_{\alpha,\beta}(\omega) = \lim_{\epsilon\rightarrow 0^+} {{e^{-\hbar\omega/k_BT}-1}\over{\hbar \omega A}}\int^\infty_0 e^{-\epsilon\tau}sin\omega\tau \\ \times 2Im\braket{\phi_2(\tau)|j_\alpha|\phi_1(\tau)}_\beta d\tau.
\end{aligned}
\end{equation}
Here, $A$ is the area of the unit cell per layer, and wave functions
\begin{equation}
\begin{aligned}
\left| \phi_1(\tau)\right>_\beta = e^{-iH\tau/\hbar}[1-f(H)]j_\beta\left| \phi_0\right>, \\
\left| \phi_2(\tau) \right> = e^{-iH\tau/\hbar} f(H)\left| \phi_0 \right>,
\end{aligned}
\end{equation}
where $f(H)=1/(e^{\beta(H-\mu)}+1)$ is the Fermi-Dirac distribution operator. 

For both density of states and optical conductivities, exponential decayed time windows are multiplied to the time-dependent correlations before performing the Fourier transform, in order to improve the approximation of the integrals. All the results are averaged with a number of different realizations of random sequences in the initial states to reduce the fluctuations appeared in the spectrum due to limit size of the sample.

\section{RESULTS AND DISCUSSION}\label{approximant}
In this paper, graphene quasicrystal is simulated by the tight-binding model based on $p_z$ orbitals, where the hopping energy between site $i$ and $j$ \cite{LCAO}
\begin{equation}
t_{ij} = n^2 V_{pp\sigma}(\left|\bm{r_{ij}}\right|)+(1-n^2)V_{pp\pi}(\left|\bm{r_{ij}}\right|).
\end{equation}
Here, $n$ is the direction cosine of relative position vector $\bm{r_{ij}}$ with respect to $\bm{e_z}$. The Slater and Koster parameters $V_{pp\sigma}$ and $V_{pp\pi}$ have the following form:
\begin{eqnarray}
V_{pp\pi}(\left|\bm{r_{ij}}\right|)=-\gamma_0 e^{2.218(b-\left|\bm{r_{ij}}\right|)}F_c(\left|\bm{r_{ij}}\right|), \\ 
V_{pp\sigma}(\left|\bm{r_{ij}}\right|)=\gamma_1 e^{2.218(h-\left|\bm{r_{ij}}\right|)}F_c(\left|\bm{r_{ij}}\right|).
\end{eqnarray}
The interlayer distance $h$ and nearest carbon-carbon distance $b$ are chosen to be 3.349 and 1.418 {\AA} respectively. $\gamma_0$ and $\gamma_1$ are turned to 3.12 and 0.48 eV to fit the experimental Fermi velocity (8.9$\sim$10.0$\times10^5$ $m/s$), respectively\cite{science_QC,pnas_QC}. $F_c$ is a smooth function
\begin{equation}
F_c(r) = (1+e^{(r-0.265)/5})^{-1}.
\end{equation}

Under a magnetic field, the hopping $t_{ij}$ will be replaced by a Peierls substitution\cite{TBPM}. 

\subsection{Approximant construction}

\begin{figure}[!htbp]
	\centering
	\includegraphics[width=8 cm]{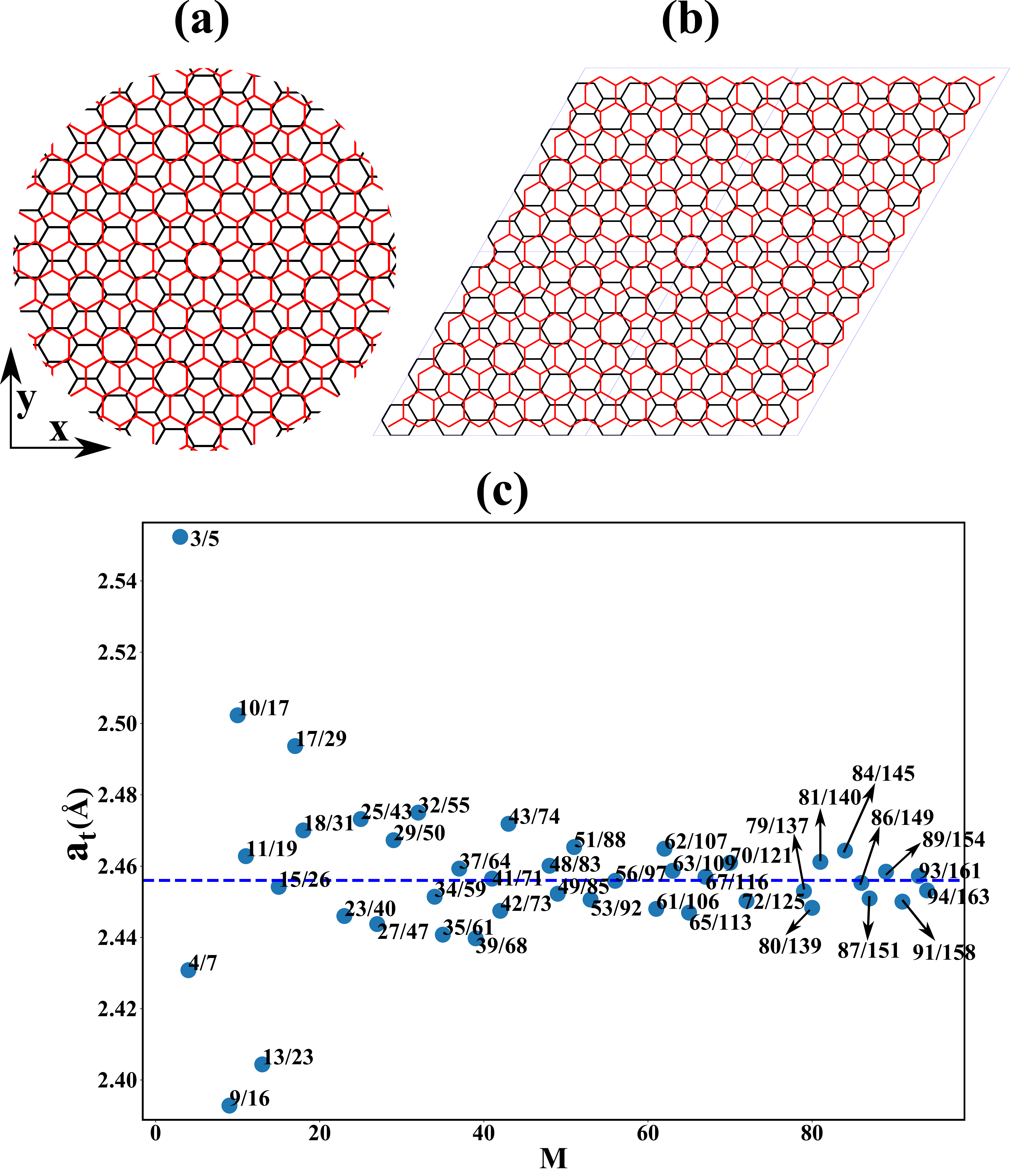}
	\caption{The atomic structures of graphene quasicrystal (a) and 4/7 approximant with four elementary unit cells (b). (c) Lattice constant $a_t$ of the top graphene layer in approximants with $M<100$. The horizontal dashed line shows the lattice constant $a=2.456$ {\AA} of pristine graphene.}
	\label{fig:struct_atop}
\end{figure}

The structure of graphene quasicrystal is shown in Fig. \ref{fig:struct_atop}(a). Along x direction, bottom layer (black) and top layer (red) have the period $3b$ and $a$ ($=\sqrt{3}b$), respectively. If a common period exists in such a structure, there should be two integers $M$ and $N$ that satisfy $M\times 3b=N\times a$, i.e., $N/M=\sqrt{3}$. Graphene quasicrystal posses the quasi-periodicity because $\sqrt{3}$ is an irrational number, namely $N/M=\sqrt{3}$ is never satisfied. The commensurate configurations of tBG with twist angle close to 30$\degree$ can be used as the approximant, but the 12-fold rotational order will be destroyed. In this paper, we construct the approximant in the following way: the twist angle of 30$\degree$ is fixed and the top graphene layer is compressed or stretched to satisfy the condition $M\times 3b=N\times a_t$, where $a_t$ is the lattice constant of the top graphene layer with strain. This method was applied to construct the periodic structure to calculate the formation energy of graphene quasicrystal by using first principle calculations\cite{pnas_QC}.

\begin{figure}[!htbp]
	\centering
	\includegraphics[width=8 cm]{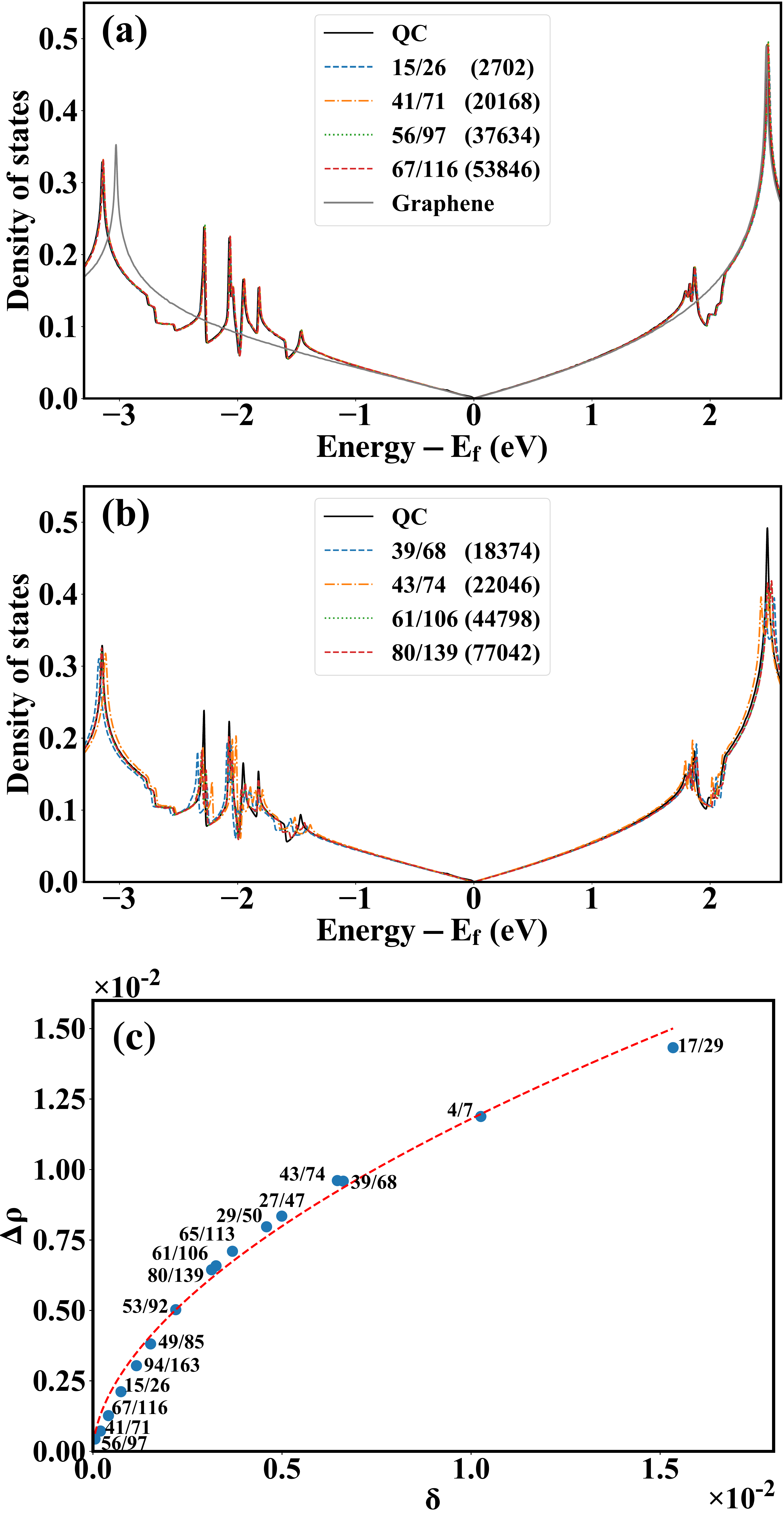}
	\caption{The comparison of DOS obtained from graphene quasicrystal (QC) and its approximants. For graphene quasicrystal, a round disk with 10021404 atoms has been used. For approximants, the numbers of carbon atoms in a unit cell are given in brackets. (a) Accurate approximants with less than 60000 carbon atoms in a unit cell, and the smallest one (15/67) contains only 2702 atoms. DOS of pristine graphene in a monolayer is also given for reference. (b) Examples of several inaccurate approximants. (c)The relative DOS differences between approximants and graphene quasicrystal vary with respective to the lattice mismatch. Red dashed line is the fitting function $\Delta \rho=0.157\times\delta ^{0.562}$.}
	\label{fig:dos}
\end{figure}

\begin{figure*}[!htbp]
\centering
\includegraphics[width=14 cm]{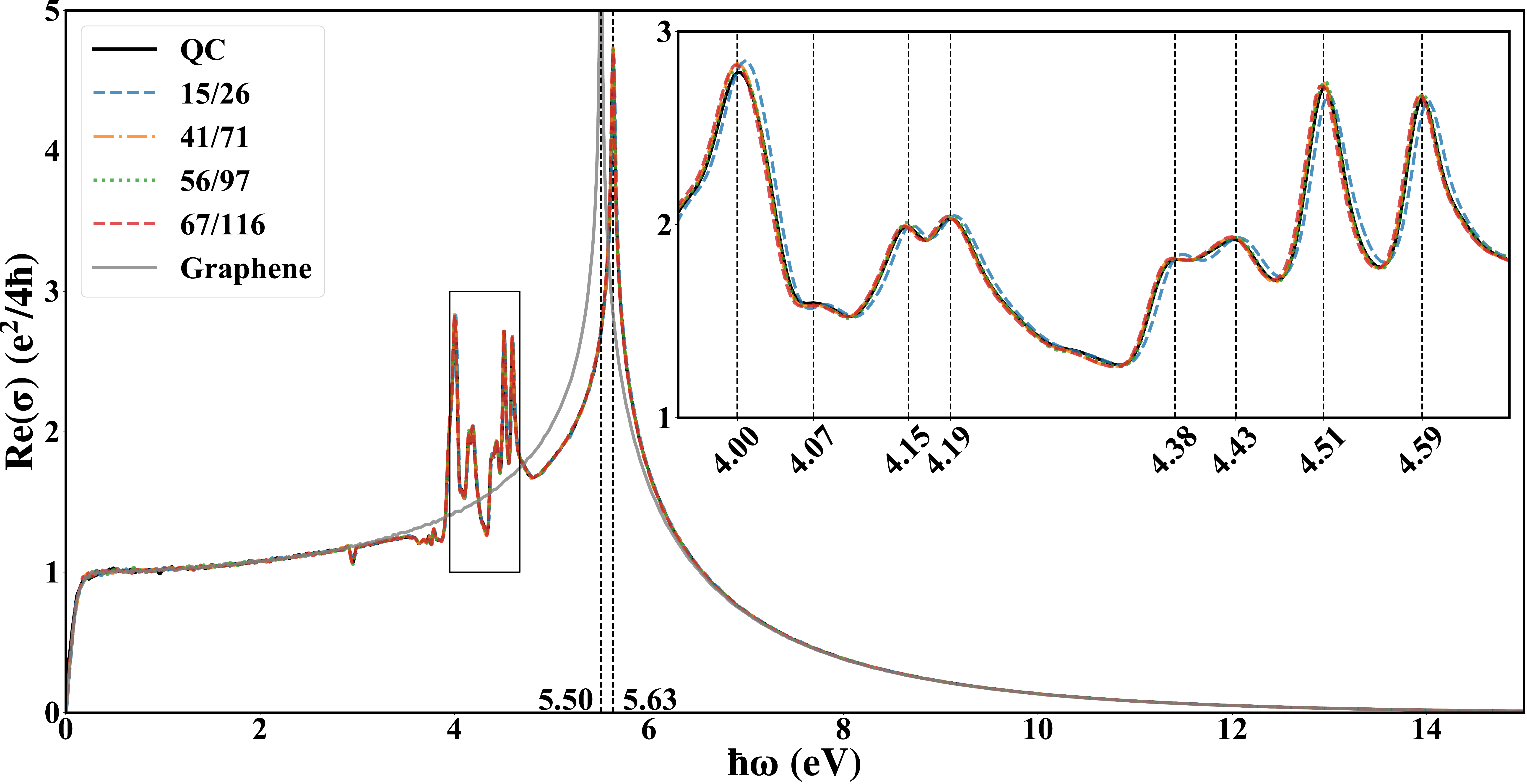}
\caption{The comparison of optical conductivities of graphene quasicrystal, the approximants and monolayer graphene. The inset shows a zoom of the peaks associated with the quasi-periodic states appeared in graphene quasicrystal, which is all accurately reproduced by proposed approximants.}
\label{fig:ac}
\end{figure*}
      
The procedure to construct the approximant of graphene quasicrystal is given below:
For a specific integer $M$ (in this paper, all integers less than 100 are considered), $N$ can be determined by
\begin{equation}
N = INT\left( {{M\times 3b}\over{a}}\right)
\end{equation}
where, $INT(x)$ stands for the integer closest to $x$. Then, $a_t$ can be obtained by 
\begin{equation}
a_t = {{M\times 3b}\over{N}}.
\end{equation}
We name such an approximant as $M/N$, which has the lattice vectors $\bm{a}_1 = (M\times 3b, 0)$ and $\bm{a}_2=({{1}\over{2}}M\times 3b, {{\sqrt{3}}\over{2}}M\times 3b)$. The structure of $4/7$ approximant is shown in Fig. \ref{fig:struct_atop}(b). Note that if $M$ and $N$ share a common divisor, or $N$ can be divisible by 3, a smaller elementary unit cell exists. Then, the approximants with such $M$ and $N$ will not be considered. The curve of the lattice constant $a_t$ with respective to the approximant size is given in Fig. \ref{fig:struct_atop}(c). It indicates that $a_t$ converges to $a$ with the increase of the approximant size. Usually, the approximant with larger size is expected to reproduce more accurately the electronic properties of the quasicrystal. However, this is not always the case as we will show in the following, and it is important to figure out what are dominant factors that determine the accuracy of these approximants shown in Fig. \ref{fig:struct_atop}(c).    

\subsection{Density of states}
The DOS of graphene quasicrystal is shown in Fig. \ref{fig:dos}(a). It is calculated from a round disk with radius about 204 nm by using the TBPM\cite{TBPM}, and is in good agreement with previous result\cite{arXiv_QC}. As shown in Fig. \ref{fig:dos}(a), comparing the DOS of graphene quasicrystal with graphene, the significant difference is the emergence of some peaks in the spectrum, which is attributed to the interaction between layers \cite{arXiv_QC}. In the vicinity of Fermi level, the DOS of graphene quasicrystal is almost the same as pristine graphene, which indicates that their electronic and optical properties may behave similar at low energies.

In order to figure out the influence of lattice constant ($a_t$) and approximant size ($M$) on the accuracy of the approximants, we compare the DOS of graphene quasicrystal with these obtained from approximants in Fig. \ref{fig:dos}(a) and (b). The results indicate that as long as $a_t$ is close enough to $a$, even a small approximant, such as 15/26 which contains only 2702 carbon atoms, can reproduce the DOS of graphene quasicrystal with very high accuracy that one can not distinguish them by eye (see the results plotted in Fig. \ref{fig:dos}(a)). However, if $a_t$ is far from $a$, even the sample contains much more atoms, and it can not be supposed to be a qualified approximant. For example, as the data shown in Fig. \ref{fig:dos}(b), 80/139 contains 77042 carbon atoms in the unit cell, but the calculated DOS differs from the one obtained from quasicrystal, especially around van Hove singularities.

The quantitative measurement on the accuracy of an approximant can be described by the standard derivate
\begin{equation}
\Delta \rho = \sqrt{{{1}\over{N}}\sum_i\left| \rho^{QC}(\epsilon_i)-\rho(\epsilon_i)\right|^2}
\end{equation}
where $N$ is the number of energy points, $\rho^{QC}(\epsilon_i)$ and $\rho(\epsilon_i)$ are the DOS of graphene quasicrystal and the approximant at energy $\epsilon_i$, respectively. We plot $\Delta \rho$ of several approximants as a function of the relative lattice mismatch $\delta$ ($=\left| a-a_t\right|/a$) in Fig. \ref{fig:dos}(c), which show clearly that $\Delta \rho$ increases monotonically with increasing lattice mismatch. It is obvious that the smaller lattice mismatch is, the more accurate DOS the approximant can reproduce. More precisely, the data can be fitted with a simple function $\Delta \rho=0.157\times\delta^{0.562}$. Our numerical tests indicate that the lattice mismatch is a crucial factor to justify the accuracy of the approximant. In the following, we will only focus on these accurate approximants shown in Fig. 2(a), such as 15/26, 41/71, 67/116 and so on.       

\subsection{Optical conductivity}
Density of states is the counting of states at certain energy, and it does not contain the details of the wave function, such as the amplitude and phase distribution in the space. To make the verification of our approximants more complete, the comparison of optical conductivities among graphene quasicrystal, its approximants and graphene is shown in Fig. \ref{fig:ac}. The calculation of optical conductivity (see the Method section for details) is based on tight-binding propagation method\cite{TBPM} without diagonalization of the Hamiltonian matrix. Although the method itself does not using directly eigenstates of the Hamiltonian, it is equivalent to the standard Kubo formulation which calculates excitations between occupied and unoccupied eigenstates. Optical conductivity is indeed a bulk property determined by wave functions, and can be used to check the accuracy of our approximants. Indeed, as we see from the results plotted in Fig. 3, there is a perfect agreement between graphene quasicrystal and its approximants with small mismatch as these shown in Fig. 2(a). Importantly, by comparing with the results of pristine graphene, there are emerged peaks around 4.0$\sim$4.6 eV in the optical spectrum of graphene quasicrystal, which are reproduced exactly with the same energies and amplitudes by proposed approximants. These peaks in the optical spectrum are attributed to Van Hove singularities appeared in density of states shown in Fig. 2(a). It is possible to identify the one-to-one correspondence of the peaks in optical spectrum and density of states by using the band structure of proposed approximant. We leave these detailed studies in future, together with other optical and plasmonic properties of graphene quasicrystals. The peak at 5.63 eV corresponds to the transition between the singularities of graphene monolayer, which shifts towards higher frequency by about 0.13 eV than graphene. 

\subsection{Eigenstates with 12-fold rotational symmetry}
\begin{figure}[!htbp]
\centering
\includegraphics[width=8 cm]{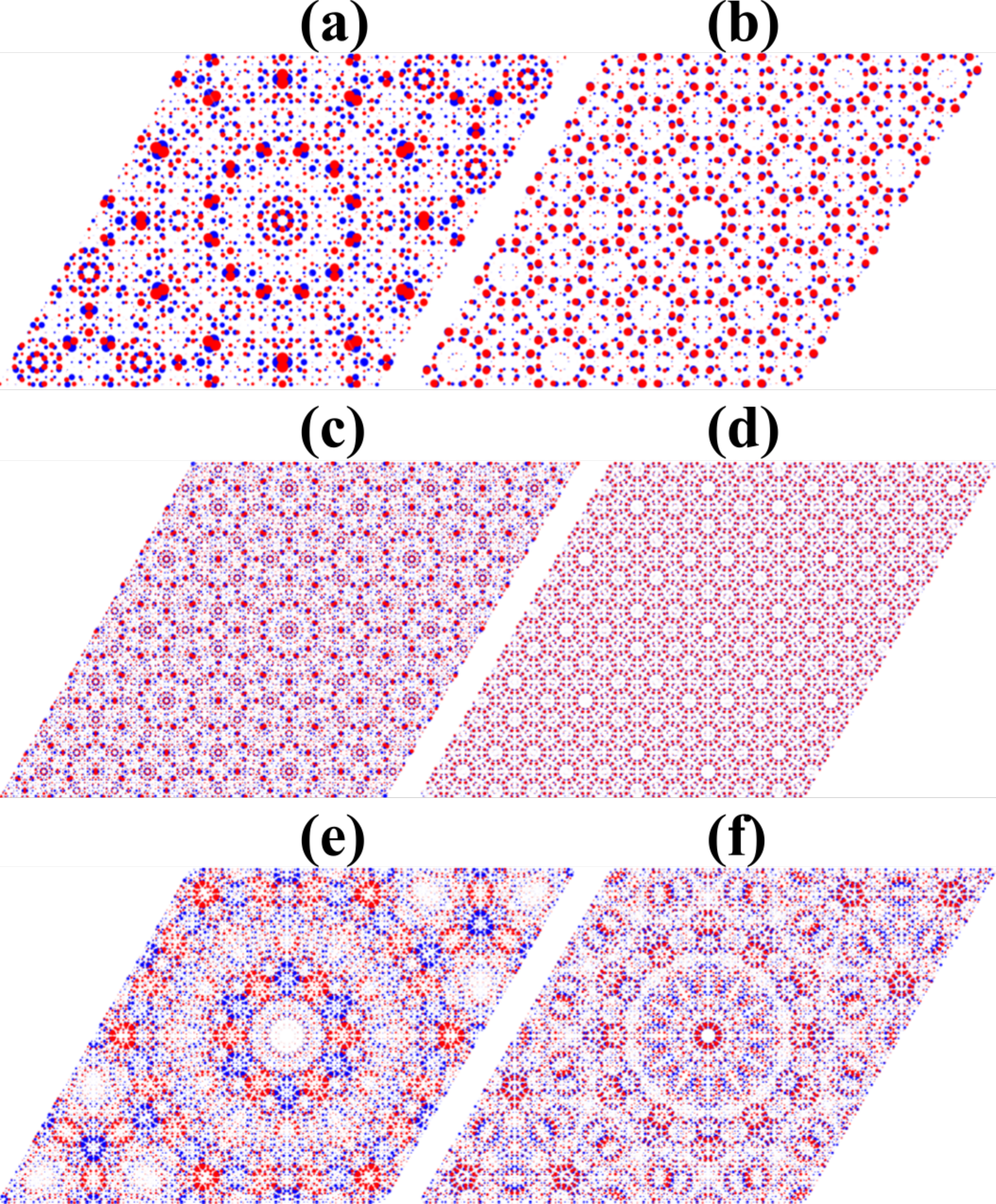}
\caption{The 12-fold rotationally symmetric eigenstates at energy (a) -4.2 and (b) -2.76 eV for 15/26 approximant as well as (c) -4.2, (d) -2.76 eV,  (e) -2.4 and (f) -2.23 eV for 41/71 approximants. Red and blue circles represent the projection on the top and bottom layers, respectively.}
\label{fig:vec}
\end{figure}

In spite of the periodicity of approximants, the quasi-periodicity still remains inside each unit cell. The electronic properties related to the quasi-periodicity can be verified by the extistance of the 12-fold rotational symmetric eigenstates within the unit cell. In this paper, 12-fold rotational symmetry means a rotation of $60\degree n + 30\degree$ followed by the mirror reflection with the mirror plane in the middle of the two graphene layers plus just rotation of $60\degree n$ ($n$ is any integer), because after these operations the structure remains unchanged. Two 12-folded rotational symmetric eigenstates of 15/26 approximant are shown in Fig. \ref{fig:vec}(a) and (b), and the corresponding states in 41/71 approximant are shown in Fig. \ref{fig:vec}(c) and (d) respectively. Comparing with 15/26 approximant, some new 12-fold rotational symmetric eigenstates appear for the 41/71 approximant, two of which are given in Fig. \ref{fig:vec}(e) and (f). Such a result is reasonable. For a real graphene quasicrystal, some critical eigenstates expand more than 20 nm in space\cite{arXiv_QC}, which can not be simulated by a small approximant, for instance, the approximant with 29.84{\degree} twist angle in Ref\cite{arXiv_QC} and the 15/26 approximant in our model. But as the approximant size increased to be large enough, these critical states appear again. 

\subsection{Effective band structure}\label{EBS}

\begin{figure*}[!htbp]
\centering
\includegraphics[width=14 cm]{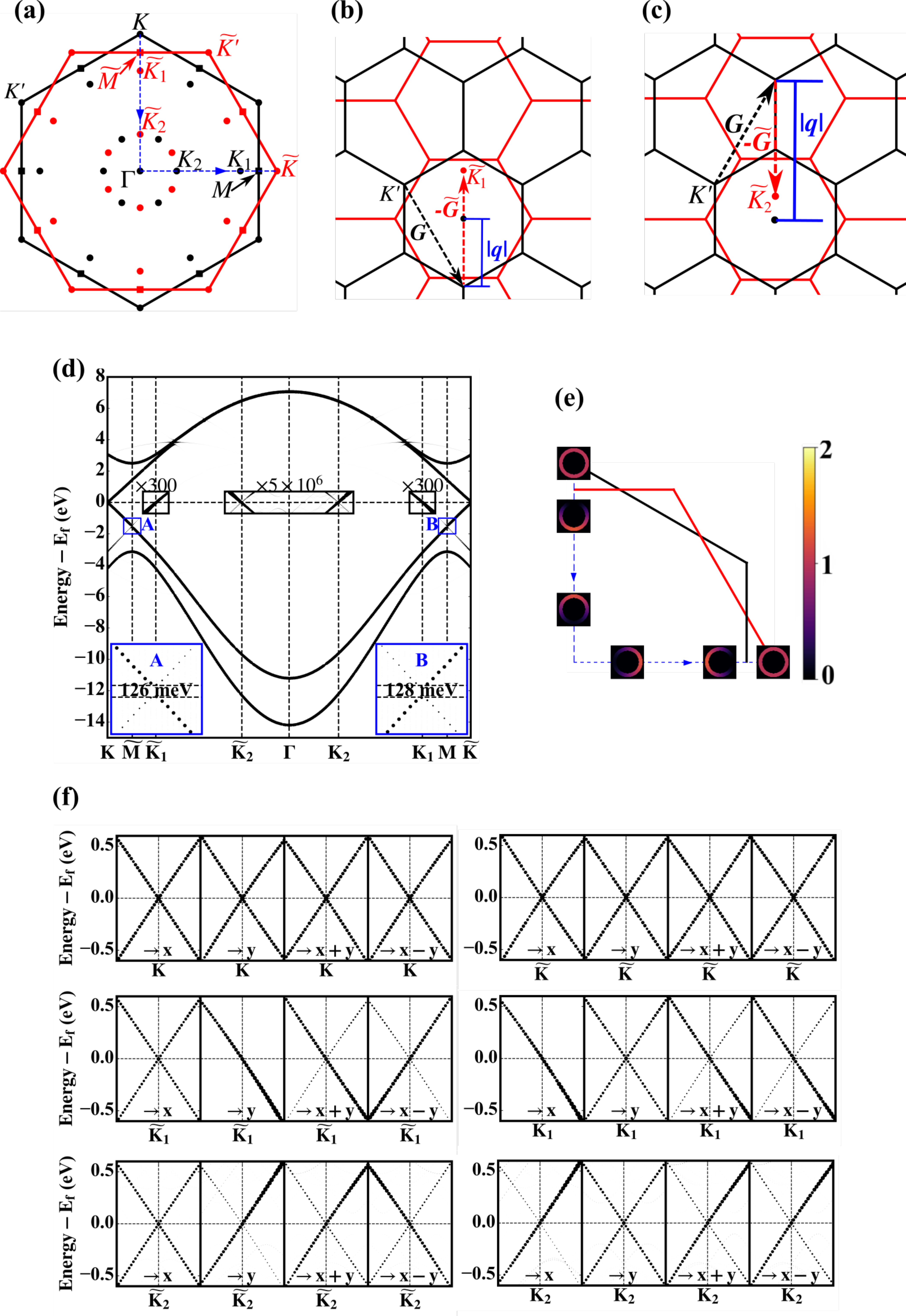}
\caption{(a) The BZs of the two graphene layers. (b) and (c) are the first two strongest scattering paths from $K^{'}$ in bottom layer to $\widetilde{K}_1$ and $\widetilde{K}_2$ in top layer respectively. (d) The EBS of 15/26 approximant. The zooms in at the bottom left and right corners show the gap at $M$ and $\widetilde{M}$ clearly. (e) The Fermi surfaces around the Dirac points above Fermi energy by 35 meV. (f) Detailed EBS around some k points along four directions. Note: In Fermi surface (e) and EBS (d) and (f), the larger spectral function is described by lighter color and larger black dot respectively. The spectral functions around Dirac points at $K_1$/$\widetilde{K}_1$ and $K_2$/$\widetilde{K}_2$ Dirac points are multiplied by 300 and $5\times10^6$ respectively.}
\label{fig:EBS_FS}
\end{figure*}

According to the generalized Umklapp scattering theory\cite{TB_U1,TB_U2}, the interlayer interaction will couple the wavevector $\bm{k}$ in the bottom layer and the wavevector $\widetilde{\bm{k}}$ in the top layer if $\bm{k}+\bm{G} = \widetilde{\bm{k}} +\widetilde{\bm{K}}$ is satisfied, where $\bm{G}$ and $\widetilde{\bm{G}}$ are the reciprocal lattice vectors of bottom and top layers respectively. The coupling matrix element $\left|\braket{\widetilde{k},\widetilde{X}|U|k,X}\right|$ can also be understood as the scattering from the $k$ state of bottom layer to $\widetilde{k}$ state of top layer, where $\left| k,X\right>$ ($\left| \widetilde{k}, \widetilde{X}\right>$) is the Bloch basis function in sublattice $X$ ($\widetilde{X}$) for bottom (top) layer, $U$ is the interlayer interaction. The scattering strength just depends on the length of $\left| \bm{q}\right|=\left| \bm{k}+\bm{G} \right|=\left| \widetilde{\bm{k}}+\widetilde{\bm{G}} \right|$, because the hopping energy between two layers is isotropic along the in-plane direction. And smaller $|\bm{q}|$ corresponds to the stronger scattering process. The Brillouin zones (BZs) of two graphene layers are shown in Fig. \ref{fig:EBS_FS}(a). If we take $K$, $K^{'}$, $\widetilde{K}$ and $\widetilde{K}^{'}$ as the original points for scattering, where the Dirac cones exist for two layers, they will be scattered to their mirrored points $K_1$ and $\widetilde{K}_1$ after the strongest scatterings and $K_2$ and $\widetilde{K}_2$ after the second strongest scatterings, respectively. As an example, the first two strongest scattering paths from $K^{'}$ are shown in Fig. \ref{fig:EBS_FS}(b) and (c) respectively.

Previous ARPES measurements\cite{science_QC,pnas_QC} and theoretical studies\cite{science_QC} show that the band structures around these six kpoints ($K$, $\widetilde{K}$, $K_1$, $\widetilde{K}_1$, $K_2$ and $\widetilde{K}_2$) are Dirac cones. The Fermi velocities for $\widetilde{K}$, $K$, $K_1$ and $\widetilde{K}_1$ Dirac cones are 9.3, 9.2, 8.9 and 9.1 $\times10^5$ m/s, respectively\cite{science_QC}. Due to the hybridization between the Dirac cones at $K$ and $\widetilde{K}_1$, a $\sim$200 meV gap can be observed at $M$ point below the Fermi level for the graphene quasicrystal on Pt(111) substrate\cite{pnas_QC}.        

Now, we check whether the approximants we proposed can reproduce the experimental results. Because the approximant contains lots of unit cells of the two layers, the band structure calculated directly from the approximant can not be used to compare with the ARPES measures. The effective band structure (EBS) derived by applying the band-unfold method can overcome this problem. Here we focus on only the smallest approximant (15/27). The EBS along the path plotted by blue dashed line in Fig. \ref{fig:EBS_FS}(a) is given in Fig. \ref{fig:EBS_FS}(d), which contains about four bands in the whole energy region. That is because only two $p_z$ orbitals exist in the unit cell of each monolayer. The strength of the spectral function becomes weaker at the Fermi level from $K$ ($\widetilde{K}$) to $\widetilde{K}_2$($K_2$) via $\widetilde{K}_1$ ($K_1$). It just follows the order of scattering strength, namely, the weaker scattering will lead to weaker strength of the spectral function of the Dirac cones at the ending points. Such results are in accordance with the generalized Umklapp scattering theroy\cite{science_QC} and the ARPES measurements\cite{pnas_QC}. In order to show the results clear, the spectral functions around $K_1$($\widetilde{K}_1$) and $K_2$ ($\widetilde{K}_2$) are always timed by 300 and 5$\times10^6$ respectively in Fig. \ref{fig:EBS_FS}. The Fermi surfaces and the detailed EBS around the six kpoints are shown in Fig. \ref{fig:EBS_FS}(e) and (f) respectively, which show the band structures are all Dirac cones clearly. Although our results indicate that the four Dirac cones at $\widetilde{K}$, $K$, $\widetilde{K}_1$ and $K_1$ have the same Fermi velocity of $9.08\times10^5$ m/s, the value is still very close to that observed in experiment. Moreover, the spectral function shows strong anisotropic intensity, which leads to the strong contrast of two bands along y direction for $\widetilde{K}_1$ and $\widetilde{K}_2$ and along x direction for $K_1$ and $K_2$, and almost half circles of the Fermi surfaces at $K_1$, $\widetilde{K}_1$, $K_2$ and $\widetilde{K}_2$. As shown in the insets in Fig. \ref{fig:EBS_FS}(d), the band gap at $M$ point below Fermi level, which has been observed in experiment\cite{pnas_QC}, can also be obtained by our model, although the gap value ($\sim$130 meV) is a little smaller than that in experiment ($\sim$200 meV). 

It is worth noting that our calculations show almost equivalent results for the two graphene layers because of their almost the same lattice constants. However, for the graphene quasicrystal grown on Pt(111) surface, mirrored Dirac points of only top layer can be detected\cite{pnas_QC}. But for the one grown on 4H-SiC(0001) surface, the mirrored Dirac points of both layers can be detected, although the ARPES signals are obviously different\cite{science_QC}. The signal is even stronger for some mirrored Dirac cones than for the original ones\cite{science_QC}. Besides, graphene quasicrystal on 4H-SiC(0001) should be n-type doped according to the ARPES measurements\cite{science_QC}. Therefore, both Pt(111) and 4H-SiC(0001) substrates should impact the electronic properties of graphene quasicrystal. The existence of substrates may be the reason why our results are not exactly the same as the experimental results. However, our theoretical results from approximant 15/27 are in accordance with generalized Umklapp scattering theory\cite{science_QC} and main experimental results\cite{science_QC}.

\subsection{Landau levels}\label{LLS}
\begin{figure}[!htbp]
\centering
\includegraphics[width=8 cm]{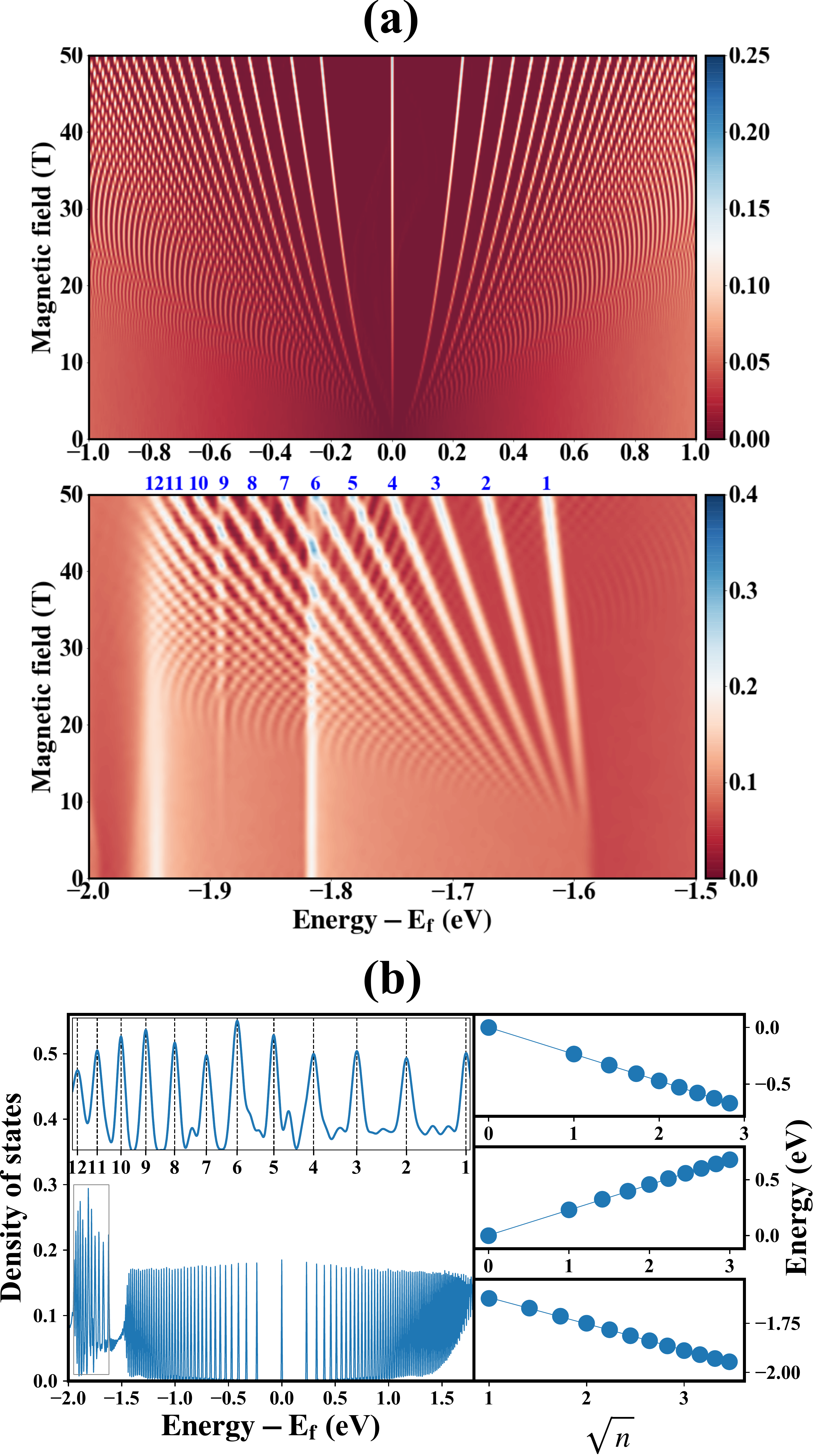}
\caption{(a) Hofstadter's butterflies of 41/71 approximant with magnetic field less than 50 T and the energy in the range of -1.0$\sim$1.0 eV and -2.0$\sim$-1.5 eV respectively. Colour bar stands for the value of DOS. The blue numbers in the bottom panel indicate the indexes of the corresponding Landau levels. (b)The DOS of 41/71 approximant (left) and Landau levels fitting (right) under a 50 T magnetic field. The inset is the zoom in of the new emerging Landau levels in the energy range about -2.0 $\sim$ -1.6 eV. For the Landau levels fitting, the top and middle panels correspond to the Landau levels of the holes and electrons, respectively, and the bottom panel for Landau levels in the inset. }
\label{fig:butterfly}
\end{figure}

For graphene under a strong magnetic field perpendicular to the graphene plane, there is a quantization of the energy states, the so-called Landau levels\cite{LL_G1,LL_G2,LL_G3,LL_G4}. The two dimensional Dirac fermion behavior of graphene can be expressed by the Landau levels which satisfies $E_n = v_F\sqrt{2e \hbar B\left|N\right|}$. In order to study the electronic properties of graphene quasicrystal under magnetic field, we calculate the Hofstadter's butterfly using 41/71 approximant under the magnetic field less than 50 T (see Fig. \ref{fig:butterfly}(a)). Similar to graphene, the DOS of the graphene quasicrystal under magnetic field also shows the Landau levels. In the vicinity of the Fermi energy, they follow the two dimensional Dirac fermion behavior but with the Fermi velocity $v_F$ of 9.20 and 8.84$\times10^5$ m/s for hole and electron, respectively, which are obtained by fitting Landau levels under 50 T magnetic field (see Fig. \ref{fig:butterfly}(b)). It agrees well with the values from the EBS (9.1$\times 10^5$ m/s) and experiments (8.9 $\sim$ 9.3$\times10^5$ m/s)\cite{science_QC}. Moreover, it is almost the same as but slightly smaller than that in graphene (9.25 and 9.05 $\times10^5$ m/s for hole and electron, respectively, which are calculated by using the same tight-binding parameters as one layer in graphene quasicrystal). It means the electronic properties of graphene quasicrystal should be similar to graphene at low energies. More interestingly, some new Landau levels appear below Fermi level by about 1.6 eV when magnetic field is more than 10 T. By fitting the new Landau levels, it can be found that they also follow the two dimensional Dirac fermion behavior but with a reduced Fermi velocity 5.21 $\times10^5$ m/s. Besides, the Landau level of n=0 doesn't exist, but its position is predicted to be around 1.49 eV below Fermi level by interpolation. It is also around the position where the band gap at $M$ appears, and the $K$($\widetilde{K}$) and $\widetilde{K}_1$($K_1$) valleys hybridize strongest.          


\section{CONCLUSION}\label{conclusioin}
In this paper, we performed a systematic study of the electronic properties of graphene quasicrystal in the framework of tight-binding approximation. Large-scale calculations of round disks of graphene quasicyrstal with more than ten million atoms have been implemented to model the real graphene quasicrystal. Furthermore, we proposed a series of approximants with translational symmetry to represent graphene quasicrystal, and the accuracies of these approximants have been verified by comparing their density of states and optical conductivities to the large round disk of graphene quasicrystal. The number of atoms in these approximants are only few thousands or tens of thousands. The lattice mismatch between two layers in the approximant is found to be the dominant factor, which determines the accuracy of this approximation. An approximant with smaller mismatch can approximate graphene quasicrystal better than the one with larger mismatch, independent on the size of the approximant. This is indeed a quite surprising result as one would expect that larger approximant would leads to better accuracy. In fact, decreasing lattice mismatch between layers should be a designing principle when building approximants for any incommensurate layered structure.

Furthermore, by applying band unfolding procedure to the smallest approximant with 2702 atoms, the effective band structure of graphene quasicrystal can be derived and compared directly with recent ARPES measurements. Such a comparison indicates that its properties agree well with the main experimental results, such as: (1) the emergence of new Dirac points, especially the mirrored ones, (2) the appearance of a band gap $\sim$130 meV ($\sim$200 meV in experiment) at the $M$ points below the Fermi energy, and (3) the Fermi velocity of $\sim 9.1\times10^5$ m/s. Besides, our results show a strong anisotropic intensity in spectral function around these new Dirac points. By calculating the density of states in the presence of strong perpendicular magnetic fields (B$>$10 T), we found that besides the usual Landau level spectrum in the vicinity of Fermi energy, which is almost the same as monolayer graphene, a group of new Landau levels appear with energies about 1.6 eV below the Fermi energy. They also follow the property of two-dimensional Dirac fermion $E_n = v_f\sqrt{2e\hbar nB}$  but with a reduced Fermi velocity of $\sim5.2\times10^5$ m/s. Interestingly, the zero order Landau level does not appear in the spectrum, no matter how strong the magnetic field is. The optical conductivities of graphene quasicrystal and its approximants have been studied numerically by using the tight-binding propagation method within the linear response theory. The optical excitations associated with quasicrystal states have been observed between 4.0 $\sim$ 4.6 eV, which should be measurable in optical experiments. 

Importantly, our proposal of the approximants for graphene quasicrystal can be applied directly for any bilayer or multilayer quasicrystals formed by two-dimensional honeycomb lattices. It works not only for graphene, but also for other materials such as hexagonal transition metal dichalcogenides, e.g., $MoS_{2}$,  $WS_{2}$,  $MoSe_{2}$ and  $WSe_{2}$, etc. In fact, to model accurately these materials, there are much more orbitals and hoppings that need to be considered in the tight-binding model\cite{cappelluti2013tight,fang2015ab}, therefore, an approximation with finite unit cell will dramatically simplify the complexities of the modelling. The numerical costs such as memory and CPU time in the numerical calculations will be reduced significantly. The studies of layered quasicrystals with other rotational order and/or other materials by using the principles proposed in this paper will be continued in future works.

\section*{ACKNOWLEDGEMENTS}
This work is supported by the National Key R$\&$D Program of China (Grant No. 2018FYA0305800) and China Postdoctoral Science Foundation (Grant No. 2018M632902). MIK acknowledges a support by the JTC-FLAGERA Project GRANSPORT. Numerical calculations presented in this paper have been performed on the supercomputing system in the Supercomputing Center of Wuhan University.

\section*{AUTHOR CONTRIBUTIONS}
G. Yu and Z. Wu performed the calculations. G. Yu wrote the manuscript. All authors analyzed the results and revised the manuscript. S. Yuan conducted the project. 

\textbf{Competing interests:} The authors declare no competing interests.

\bibliography{30_tBG_QC}
\end{document}